\documentclass[twocolumn,showpacs,twoside,10pt,pra,superscriptaddress]{revtex4-2}
\usepackage{graphicx}
\usepackage{dcolumn}
\usepackage{bm}
\usepackage{braket}
\usepackage{amsmath}
\usepackage{hyperref}
\usepackage[T1]{fontenc}
\renewcommand{\d}{\mathrm{d}}

\begin{document}
 
\preprint{APS/123-QED}

\title{Investigation of Rare-Earth Ion-Photon Interaction and Strong Coupling in Optical Microcavities}

\author
{Quanshen Shen$^{1,2}$,
Wentao Ji$^{1,2}$,
Junyu Guan$^{1,2}$, 
Li Qian$^{1,2}$,\\
Zihua Chai$^{1,2}$,
ChangKui Duan$^{1,2,3}$,
Ya Wang$^{1,2,3}$,
Kangwei Xia$^{1,2,3*}$
\\
\normalsize{$^{1}$ Laboratory of Spin Magnetic Resonance, School of Physical Sciences,}\\
\normalsize{University of Science and Technology of China, Hefei 230026, China.}\\
\normalsize{$^{2}$ Anhui Province Key Laboratory of Scientific Instrument Development and Application,}\\
\normalsize{University of Science and Technology of China, Hefei 230026, China.}\\
\normalsize{$^{3}$ Hefei National Laboratory, University of Science and Technology of China, Hefei 230088, China.}\\
\normalsize{$^*$ E-mail: kangweixia@ustc.edu.cn}
}
 



\begin{abstract}

The strong coupling between an emitter and a cavity is significant for advancing quantum networks. 
Due to their long optical and spin coherence times, rare-earth ions (REIs) represent a compelling platform for quantum networks. 
However, their inherently weak intra-4f optical transitions typically result in low coupling strength, thus restricting most current achievements to the weak coupling regime. This work proposes a scheme to realize an on-chip quantum network by coupling REIs to high-quality whispering gallery mode (WGM) microcavities. Additionally, we provide numerical validation for a parametric amplification technique to enhance the emitter-cavity coupling strength. As an extension of this approach, the coupled system efficiently achieves the quantum entanglement of local and flying qubits. 
This study deepens the understanding of emitter–cavity interactions and contributes to realizing REIs-based photonic platforms, which are crucial to distributed quantum computing and developing robust quantum networks.

\end{abstract}

\maketitle 

\section{\label{sec:level1}Introduction }
Long-distance quantum networks are essential for quantum communications and distributed quantum computing~\cite{kimble2008quantum}, where quantum nodes can be implemented through single-emitter-based quantum systems that deterministically interface flying and local qubits.
However, the distance between quantum nodes is limited as flying qubits, typically photons, decline exponentially with distance. This challenge can be addressed using quantum repeaters~\cite {briegel1998quantum,azuma2023}. 
Solid-state spin qubits such as quantum dots~\cite{yu2023telecom_quantum_dots} and defects in solids~\cite{meng2024solid} are promising candidates for the implementation of quantum repeaters due to their abundance of optical transitions, long coherence times, and spin-photon interfaces~\cite{kinos2021roadmap,wolfowicz2021quantum}. 
In solid-state systems, rare-earth ions (REIs) exhibit long spin coherence times due to the shielding of inner 4f-4f electrons by the outer two layers of electrons. However, the 4f-4f optical transitions of rare earth ions (REIs) are dipole-forbidden, which leads to weak ion-photon interactions and low photon detection efficiency.

Incorporating REIs into photonic cavities can enhance ion-photon interactions.
When the coupling strength between the cavity and ion is lower than their decay rates, the REI's optical transition lifetime will be reduced, and its fluorescence intensity will be enhanced due to the Purcell effect~\cite{gonzalez2024light}, enabling single-ion detection~\cite{casabone2021dynamic, dibos2018atomic, zhong2018optically}. 
Experiments have demonstrated cavity-enhanced fluorescence in several REI systems, including Er:CaWO$_4$, Yb:Y$_2$SiO$_5$, Er:LiNbO$_3$, Er:Si, Er:TiO$_2$, Er:Y$_2$SiO$_5$, Yb:YVO$_4$, Yb:LiNbO$_3$~\cite{casabone2021dynamic,dibos2018atomic,zhong2018optically,ourari2023indistinguishable,ruskuc2024scalable,yang2023controlling_Er_LiNO_3,sullivan2023quasi_Er_TiO2,merkel2020coherent_Er_YSO,kindem2020control_Yb_YVO4,xia2022tunable}. 
In cases where the coupling strength surpasses the decay rates, known as the strong coupling regime, quantum information can be reversibly exchanged between photons and ions~\cite{mcauslan2009strong}, facilitating a deterministic photon-ion interface. 
Although extensive research has been conducted in other systems~\cite{reithmaier2004strong,do2024room,albrecht2013coupling}, strong coupling between REIs and optical cavities remains elusive in current experimental schemes.
 
Weak ion-photon interactions and the long optical lifetimes of REIs require both spatial and temporal confinement of light to achieve strong coupling, which presents a challenge with existing technology. 
Recent advancements have primarily been made with photonic crystal (PC) cavities~\cite{ourari2023indistinguishable, yang2023controlling_Er_LiNO_3, dibos2018atomic, zhong2018optically} and fiber-based Fabry–Pérot (FP)~\cite{casabone2021dynamic}cavities.PC cavities are particularly effective at spatial confinement, while FP cavities excel in temporal confinement. Notably, a study by D. L. McAuslan~\cite{mcauslan2009strong} has shown the potential to achieve the ''bad cavity'' strong coupling regime with ions in Whispering-Gallery Mode (WGM) cavities. WGM cavities, while maintaining a small mode volume, possess extremely high-quality factors, which may make them stand out in achieving strong coupling~\cite{duan2004}.

REIs doped into various crystals preserve their optical and spin properties, providing a wide range of host materials. For large-scale integrated quantum devices, sophisticated fabrication techniques and tunability are critical.
Lithium niobate, known for its exceptional electro-optic, piezoelectric, and nonlinear properties, offers well-established fabrication capabilities for photonic devices~\cite{zhu2021integrated}. 
Recent work has demonstrated the use of on-chip lithium niobate photonic devices to match ionic energy levels through electro-optic tuning.~\cite{xia2022tunable,yang2023controlling_Er_LiNO_3}.

In this work, we investigate a coupled emitter-cavity system based on REIs doped in microcavities, focusing on the most effective approach for realizing strong coupling between the emitter and cavity.
The results show that the WGM can significantly optimize the existing coupling parameters to achieve a Purcell factor of $10^4$. 
By leveraging the nonlinear effects in lithium niobate crystals, the interaction can be effectively amplified to achieve strong coupling.
Finally, through numerical methods, we predict that using a WGM cavity can optimize the fidelity of the existing single-photon and single-ion interaction to 0.95. 
Our work broadens the scope of existing research on rare-earth ion-cavity systems and contributes to the advancement of quantum networks based on REIs.

\begin{figure}[b]

\includegraphics[width=0.45\textwidth]{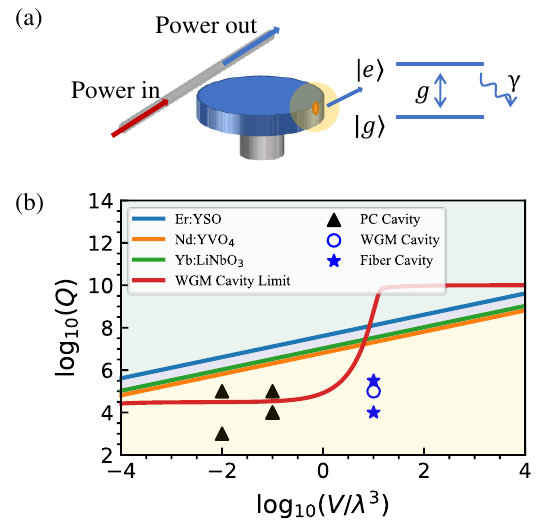}
\caption{(a) Setup and coupling scheme. A rare-earth ion with a spontaneous 
emission rate $\gamma$ is coupled with the microdisk with a coupling strength $g$. 
Excitation and readout are performed through a tapered fiber. (b) 
$Q$ factors and model volumes for different cavities. 
The red line shows the limit of the $Q$ factor for the WGM  
cavity at different model volumes. The blue (green, orange)  line  
shows the model volumes and $Q$ factors required to achieve strong 
coupling conditions for Er:Y$_2$SiO$_5$ (Yb:LiNbO$_3$, Nd:YVO$_4$). The black triangle (blue circle, blue asterisk) represents the parameter ranges achieved by PC cavities (WGM cavities, Fiber-based FP cavities).  }
\label{fig:model}
    
\end{figure} 

\section{\label{sec:level2}Modeling of the coupled system}
The interaction between REIs and an optical cavity is modeled as the interaction between a single-mode optical field and a two-level system. 
Three key parameters govern the system's dynamics: the coupling strength $g$ between a single ion and a single photon, the cavity decay rate $\kappa$, and the ion's spontaneous emission rate $\gamma$.
The Jaynes-Cummings (JC) model is applied to describe the coupled system under the rotating-wave approximation, and the Hamiltonian is given by:
\begin{equation}
    H = \hbar \omega_a \sigma^+ \sigma^- + \hbar \omega_c a^\dagger a + \hbar g (\sigma^\dagger a + \mathrm{h.c.})
    \label{eq:Hamiltonian}
\end{equation}
where $\omega_a\ (\omega_c )$ denotes the frequency of ion (cavity), $a\ (a^\dagger)$ represents the annihilation (creation) operator of the photon, and $\sigma^+\ (\sigma^-)$ corresponds to the excitation (de-excitation) operator. $\ket{e}$ is the excited state of REI and $\ket{g}$ is the ground state. 
The dynamics of the system are described by the master equation~\cite{scully1997quantum}:

\begin{gather}
    \frac{\d \rho }{\d t} = \frac{1}{i\hbar} [\rho ,H]  + (L_1 + L_2 + L_3)\rho\\
    L_1\rho = \frac{\kappa}{2} (2a^\dagger \rho a  - a^\dagger a \rho  - \rho a^\dagger a )\\
    L_2\rho = \frac{\gamma}{2} (2\sigma^+ \rho \sigma^-  - \sigma^+ \sigma^- \rho  - \rho \sigma^+ \sigma^- )\\
    L_3\rho = \frac{\gamma_p}{2} (\sigma_z\rho \sigma_z  -  \sigma_z\rho  - \rho\sigma_z )
    \label{eq:loss }
\end{gather}
where $L_1$, $L_2$, and $L_3$ are superoperators representing the cavity loss, ion spontaneous emission, and ion dephasing (with the rate $\gamma_p$), respectively. $\rho$ is the system density matrix. 

Our primary focus is the strong coupling regime, defined by $g \gg (\kappa,\gamma)$, which means we require the atom-cavity coupling to be significant while the atomic and cavity decay rates are small. This requires us to understand how these three rates depend on the  REIs and the optical microcavity.
The cavity decay rate is given by:
\begin{equation}
    \kappa = \frac{\omega_c}{2Q}
    \label{eq:decay of cavity}
\end{equation}
where $Q$ is the quality factor of the cavity. 
The coupling strength is  given by~\cite{B1}:
\begin{equation}
    g = \frac{\mu}{n} \sqrt{\frac{\omega_a}{2\hbar \epsilon_o V}}
    \label{eq:coupling strength}
\end{equation}
where $n$ is the refractive index of the cavity, $\mu$ corresponds to the dipole transition strength, and $V $ is the mode volume of the cavity.
The transition dipole moment of the atom is calculated from the spontaneous emission lifetime of the atom as~\cite{scully1997quantum}:
\begin{equation}
    \mu  = \sqrt{\frac{3 \epsilon_0 \hbar \lambda^3  }{ 8 \pi n \chi_l T_{\rm{spon}}}}
    \label{eq:lifetime}
\end{equation}
where the factor $\chi_l = (n^2 +2 )^2/ 9 $ is a local correction to the electric field to account for the fact that the ion is less polarized than the bulk medium.

For REIs, the spontaneous emission rate ($\gamma$) is typically kHz, significantly lower than the cavity decay rates in the MHz range. As a result, the strong coupling condition is achieved when $g\gg\kappa$. From Eqs.~(\ref{eq:decay of cavity}), (\ref{eq:coupling strength}), and (\ref{eq:lifetime}), it is necessary to satisfy:
\begin{equation}
    \frac{3Q^2}{16\pi^2 \chi_l k \omega T_{\mathrm{spon}}}\gg1
    \label{eq:strong coupling condition}
\end{equation}
where $k$ is defined as $ V/(\lambda^3)$. This ratio represents normalized mode volume. 
This expression indicates that achieving strong coupling for REIs requires a cavity with a high-quality factor and a small mode volume, which are contradictory in practice. Increasing the quality factor is more effective than decreasing the mode volume due to the quadratic dependence on $Q$. This highlights the importance of using ultra-high-$Q$ microcavities with moderate mode volumes. 

\section{\label{sec:level3}Simulation result  }

\begin{table*}
    \caption{\label{tab:table3}Parameters of REIs coupled with microcavities.}
    \begin{ruledtabular}
    \begin{tabular}{ccccccccc}
     Ion&$\lambda$&$\kappa/2\pi$&$g/2\pi$&$Q$ & Cavity&Purcell factor& Reference&\\ \hline
    Yb:YVO$_4$     &	984.5 nm   & 31.4 GHz & 23.9 MHz  &	$9.7\times10^3$   &PC cavity  &	27    &\cite{casabone2021dynamic}\\
    Er:LiNO$_3$    &	1533.27 nm & 2.0 GHz  &	2.8 MHz   &	$1\times10^5$     &PC cavity & 177   &\cite{yang2023controlling_Er_LiNO_3}\\
    Er:CaWO$_4$  &	1532.6 nm  & 1.0 GHz  & 2.3 MHz	  & $1.9\times10^5$   &PC cavity     & 850   &\cite{ourari2023indistinguishable}\\
    Er:MO$_2$      &	1532 nm    & 3.14 GHz & 2.49 MHz  & $6.2\times10^4$   &PC cavity     & 1040  &\cite{horvath2023strong}\\

    Er:Y$_2$SiO$_5$  &	1536 nm    & 1.25 GHz &	0.74 MHz  & $2\times10^5$   & Fiber-based FP cavity 	  & 170   &\cite{merkel2020coherent_Er_YSO}\\
    Yb:LiNO$_3$    &	980 nm     & 3.83 GHz &	1.9 MHz   & $8\times10^4$    &WGM cavity & 10    &\cite{xia2022tunable}	\\

    \end{tabular}
    \end{ruledtabular}
\end{table*}
The conditions for strong coupling differ among various rare earth ions (REIs), influenced by factors such as the ion’s emission wavelength, radiation lifetime, and the host material’s refractive index.
To find a suitable system for strong coupling, we present several cavity parameters from existing studies in FIG.~\ref{fig:model}(b) and TABLE.~\ref{tab:table3}~\cite{ourari2023indistinguishable, ruskuc2024scalable, yang2023controlling_Er_LiNO_3, sullivan2023quasi_Er_TiO2, merkel2020coherent_Er_YSO, kindem2020control_Yb_YVO4, xia2022tunable,horvath2023strong}.
The PC cavity can reach $k = 10^{-1}-10^{-2}$, and the maximum $Q$ factor is no more than $10^5$. Fiber-based FP cavities have $k \approx 10$ and $Q = 10^6$. Theoretically, WGM cavities can achieve ultra-high quality factors while maintaining specific mode volumes. Experimentally, lithium niobate-based WGM cavities with quality factors as high as $10^7$ have been achieved~\cite{zhang2019fabrication}.

The dependence of the WGM mode volume and $Q$ factor is examined and shown as the blue dashed line in FIG.~\ref{fig:model}(b). The $Q$ factor consists of three main components~\cite{righini2011whispering}, corresponding to the radiation loss (${1}/{Q_r}$), the material loss (${1}/{Q_m}$), and the scattering loss (${1}/{Q_s}$), respectively, as given by:
\begin{equation}
    \frac{1}{Q} = \frac{1}{Q_r} + \frac{1}{Q_m} + \frac{1}{Q_s}.
    \label{Q_factor}
\end{equation}
When the cavity's mode volume is comparable to the wavelength, it fails to confine the optical field effectively. 
In such cases, the radiation loss  becomes the primary source of energy dissipation.
The radiation loss decreases exponentially as the mode volume increases. When the mode volume significantly exceeds the wavelength scale, the radiation loss becomes negligible, and the primary loss is due to the intrinsic material absorption.
The third component is the scattering loss, which results from surface irregularities introduced during fabrication. The scattering loss can be substantially minimized through techniques such as chemical-mechanical polishing (CMP) or by optimizing the design of the cavity mode.
Consequently, WGM optical cavities with a high $Q$ factor approaching the intrinsic material loss have been experimentally realized, with a reported $Q$ factor of $4 \times 10^7$ and mode volume of $40~\rm{\mu m^3}$~\cite{zhang2019fabrication}.

\begin{figure}[b]

    \includegraphics[width=0.45\textwidth]{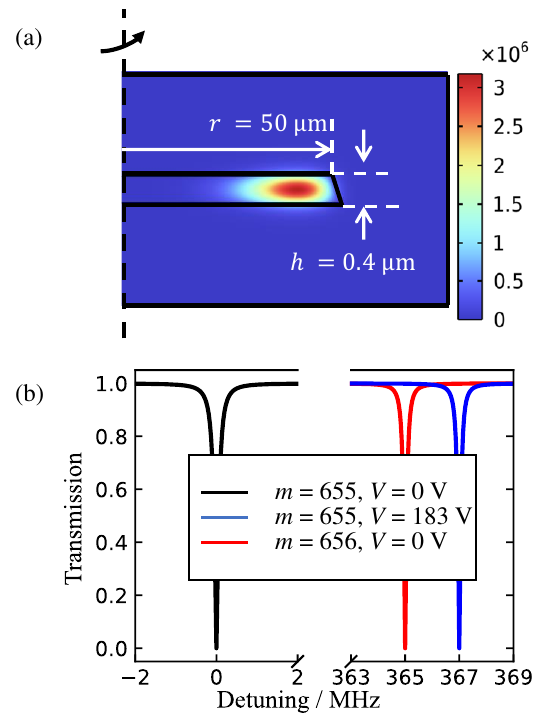}
    \caption{Simulation results of the WGM cavity (a) Electric field distribution of the TE mode of the cavity. (b) Theoretically, the transmission spectral lines of the cavity, represented by the black and red lines, correspond to two angular mode numbers that differ by one. By applying a voltage of 183 V along the Z-axis, the spectral line with mode number \( m = 654 \) can be adjusted to intersect with the spectral line of \( m = 655 \).}
    \label{fig:cavity model}
\end{figure} 

The 4f-4f transition of REIs in crystals is enabled by crystal-field-induced electric dipole moments, whose strength varies with the specific REIs and host crystals. Typical values for rare-earth 4f-4f electric dipole dipole moment range from $10^{-32}$ to $10^{-31}~\rm{C\cdot m}$.
The doping of $\mathrm{YVO_4}$ with $\rm{Yb^{3+}}$ results in a strong dipole moment of $1.83\times 10^{-31}~\rm{C \cdot m}$~\cite{kindem2018characterization},
while doping $\rm{Y_2SiO_5}$ with $\rm{Er^{3+}}$ produces a relatively weak dipole moment of $2.07\times 10^{-32}~\rm{C \cdot m}$~\cite{mcauslan2009strong}. 
This difference necessitates substantially different experimental parameters to achieve strong coupling. For a given mode volume, the $Q$ factor required to reach strong coupling is higher for Er:Y$_2$SiO$_5$ than for Yb:YVO$_4$.

Considering experimental implementations, we use the $\rm{Yb^{3+}}$ ions doped in the lithium niobate, which has a significant dipole strength of $5.7\times 10^{-32}~\rm{C\cdot m}$~\cite{B1}. 
High-energy, low-dose ion implantation precisely positions the Yb ions within the lithium niobate film, ensuring they remain in cavity modes at the maximum electric field~\cite{xia2022tunable}.
The lithium niobate WGM cavities can be fabricated through the CMP process, minimizing surface scattering losses~\cite{zhang2019fabrication}.
The simulated spectrum of the WGM cavities is shown in FIG.~\ref{fig:cavity model}(b). 
The cavity radius and the number of angular modes are adjusted to align with the ion frequency. 
For a WGM cavity with a radius of 50 $\mu$m and a thickness of $400~\rm{nm}$, a resonance peak is expected at 980 nm, with a free spectral range (FSR) of 0.365 GHz. 
Electro-optical tuning can be achieved by applying a voltage in the Z-direction using SU-8 adhesive spin-coated onto the cavity, enabling a tuning rate of $10~\rm{pm/V}$~\cite{xia2022tunable}. With an applied voltage of 180 V, the resonance frequency is shifted by 0.367 GHz, exceeding the FSR and enabling the alignment of arbitrary ion and cavity frequencies.
According to the simulations, SU-8 minimally affects the effective refractive index of the lithium niobate layer, supporting experimental ease and cavity design flexibility. 

With a  $50~\rm{\mu m}$ microdisk radius, radiation loss ($1/Q_r$) is $10^{-12}$, significantly lower than intrinsic material loss. Through the processing techniques~\cite{zhang2019fabrication} developed by Cheng Ya's team for lithium niobate microdisk cavities, a cavity decay rate $\kappa = 50~\mathrm{MHz}$ is achieved. The mode Volume of this cavity is $50~\rm{\mu m^3}$, leading to a coupling rate equal to $20~\mathrm{MHz}$.  The decay rate of the $\rm{Yb^{3+}}$ ion in the niobate microdisk is lower than $1~\rm{kHz}$.
For our simulation, the cQED (cavity Quantum electro dynamics) parameters are estimated as $(\kappa,\gamma,g) = 
(50~\rm{MHz},1~\rm{kHz},20~\rm{MHz})$, and the Purcell factor is $4 \times 10^4$.

\begin{figure}[b]

    \includegraphics[width=0.45\textwidth]{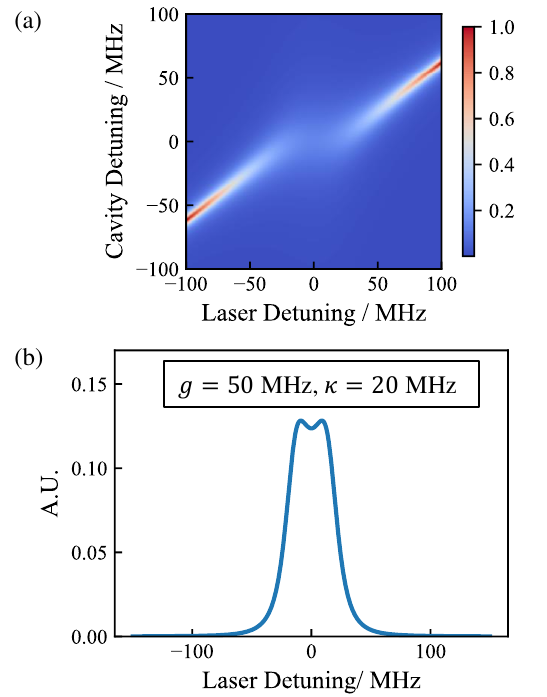}
        
    \caption{The spectrum of a Yb ion coupled by WGM cavity. (a) The spectrum of the coupled system by scanning the laser detuning and the cavity detuning at $(\gamma,\kappa,g)=(1\rm{kHz},50\rm{MHz},20\rm{MHz})$. (b)
    The spectrum of the coupled system by scanning the laser detuning for $g/\kappa = 0.5$.}
    \label{fig:spectroscopy}
\end{figure}

To characterize the system, we deduce the spectroscopy of the system through the equation~\cite{wang2021defect}:
\begin{equation}
    A_x (\hbar\omega)=C\sum_{l=1}^N\delta(\omega-\omega_l)h\omega_l C_{il}^{el}\mu^2.
    \label{eq:spectroscopy}
\end{equation}
shown as FIG.~\ref{fig:spectroscopy}(a). Here, $\omega_l$ is the frequency of the cavity mode at $l$, and $C_{il}^{el}$ is the projection of the eigenstate of the system onto the state $\ket{g,1_l}$. 
$\ket{g,1_l}$ represents the cavity model $l$ having one photon. In these parameters,
the Rabi vacuum slipping is small. When the cavity and the ion frequency are the same, the spectrum of the ion is shown in FIG.~\ref{fig:spectroscopy}(b).

\section{\label{sec:level4}amplification of coupling strength}
 
\begin{figure}[b]

    \includegraphics[width=0.45\textwidth]{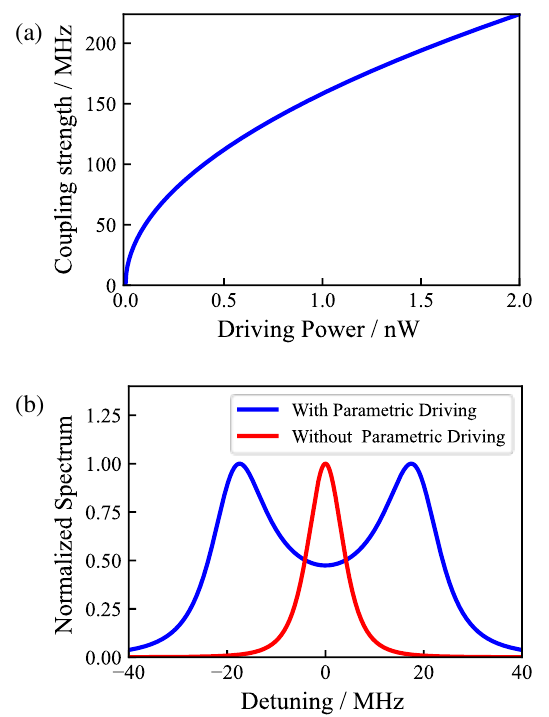}
        
    \caption{Simulation results of parametric amplification. (a) Coupling strength as a function of driving power. (b) The spectrum of the coupled system  with parametric driving(blue solid line) and without parametric driving(red solid line) 
    }
    \label{fig:parameter amplification}
        
\end{figure}

The 4f-4f dipole-forbidden optical transitions of REIs exhibit weak oscillator strength, making it difficult to achieve a coupling strength that exceeds the cavity decay rate.
However, increasing the number of coupling channels significantly enhances the coupling strength~\cite{liu2024tunable,xiao2012strongly}. Qin proposed using parametric amplification in bosonic modes to strengthen the interaction~\cite{qin2018}. We have numerically verified the feasibility of parametric amplification in a Yb-doped $\rm{LiNbO_3}$ system utilizing a WGM cavity. 
The parametric driving method modifies the medium of light-matter interaction, shifting it from vacuum to compressed vacuum, thereby increasing the interaction strength. 
The generation of single-photon-level compressed light has been experimentally 
demonstrated through periodic modulation of lithium niobate~\cite{lu2020toward}. By periodically modulating an REI-doped WGM cavity and pumping at twice the ion resonance frequency, we can derive the Hamiltonian of the system as follows~\cite{qin2024quantum}:
\begin{equation}
    \widehat{H}_{\mathrm{M}}=g(\hat{\sigma}^{\dagger}\hat{a}+\hat{\sigma}\hat{a}^{\dagger})+\Delta_{a}\sigma^{\dagger}\sigma+\Delta_{c}\hat{a}^{\dagger}\hat{a}+\frac{\Omega}{2}(\hat{a}^{2}+\hat{a}^{\dagger2}\mathrm{e}^{-i2\theta})
    \label{eq:Hamiltonian of Parametric amplification}
\end{equation}
$\Delta_a = \omega_a - \omega $, $\Delta_c = \omega_c - \omega$, $2\omega$ is the pumping frequency. A term $\Omega/2 (a^{2}+a^{\dagger2}\exp^{-i2\theta})$  is introduced through the second-order nonlinearity.
The nonlinearity of lithium niobate requires the optical cavity to support both the ion's frequency and its two-fold mode. Here,
$\Omega$ represents the driving strength, which exhibits power dependence, as described by the following formula:
\begin{equation}
    \Omega=\frac{2\sqrt{\hbar \omega_0 ^3}}{\sqrt{\varepsilon_0 V}}\chi^{(2)}\langle b\rangle 
    \label{eq:power depend} 
\end{equation}
Here, $\langle b\rangle  $ corresponds to the average number of pumping photons, and  $\chi^{(2)}$ represents the second-order nonlinear coefficient. The simulation results are shown in FIG.~\ref{fig:parameter amplification}(a). 
When pumping with a frequency-doubled laser at 1 nW power, a 10-fold amplification of coupling strength can be achieved, reaching the strong coupling regime. This manifests as distinct Rabi splitting observable in the spectrum, as illustrated in FIG.~\ref{fig:parameter amplification}(b). 

Through the Bogoliubov squeezing transformation, the annihilation operator 
$\alpha$ is expressed as:
\begin{equation}
    \alpha=a\cosh(r)+a^\dagger \exp(i\theta)\sinh(r) 
    \label{eq:BogoLiubov transformation}
\end{equation}
where \( r = \frac{1}{4} \ln \left( \frac{\Delta_c + \Omega}{\Delta_c - \Omega} \right) \)
We diagonalize the second-order terms with the final Hamiltonian expressed as:
\begin{equation}
    \begin{split}
H=&\Delta_\alpha\alpha^\dagger\alpha+g\operatorname{cosh}(r)\big(\sigma^\dagger\alpha+\sigma\alpha^\dagger\big)+\\
&g\operatorname{sinh}(r)\operatorname{exp}(-i2\theta)\big(\sigma\alpha+\sigma^\dagger\alpha^\dagger\big)+\Delta_\alpha\sigma^\dagger\sigma 
\label{eq:master equation of Parameters amplification}
\end{split}
\end{equation}
where $g \cosh(r)( \sigma^\dagger \alpha + \sigma \alpha^\dagger )$ and $g \sinh(r) \exp(-i 2 \theta) (\sigma \alpha + \sigma^\dagger \alpha^\dagger )$
  characterize the strengths of the rotating-wave and counter-rotating interactions. The counter-rotating interactions are ignored in the rotating-wave approximation. The interaction Hamiltonian in Eq.~(\ref{eq:master equation of Parameters amplification}) is approximated written as:

\begin{equation}
    \begin{split}
    H=&\Delta_\alpha\alpha^\dagger\alpha+g\operatorname{cosh}(r)\big(\sigma^\dagger\alpha+\sigma\alpha^\dagger\big) +\\
     &+\Delta_\alpha\sigma^\dagger\sigma 
    \label{eq:Hamiltonian with rotating approximation}
    \end{split}
\end{equation}
The eigenstates of the cavity mode transition from a Fock state to a compressed Fock state, causing the coupling strength to be modified by a squeezing factor $\cosh r$.

The parametric amplification method is versatile and can be applied to various materials with strong second-order nonlinearities, effectively enhancing coupling strength. However, this enhancement comes at the cost of increased noise, which can be mitigated by pumping with squeezed light ~\cite{qin2018}.
Demonstrating parametric amplification in solid-state systems, particularly at the single-photon level with squeezed light, remains challenging. Due to its strong second-order nonlinearity, periodically poled lithium niobate (PPLN)-based WGM cavities are currently among the most promising platforms for single-photon squeezed light generation~\cite{lu2020toward}.

\begin{figure}[b]
\includegraphics[width=0.45\textwidth]{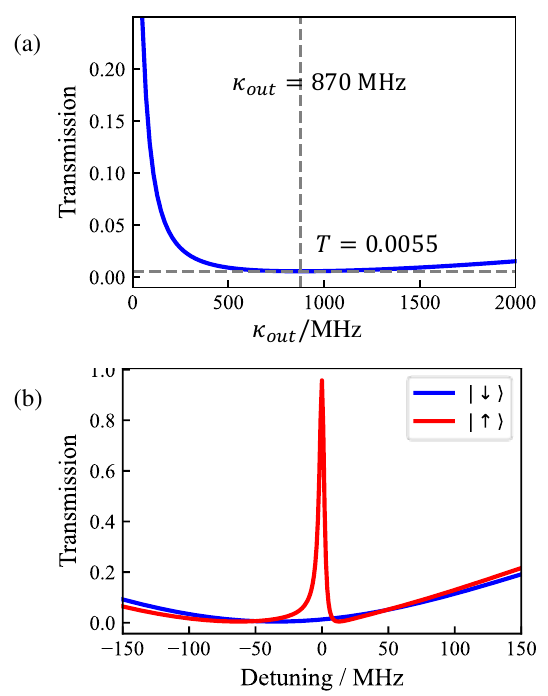}
\caption{The transmission spectrum of the coupled system. (a) Back-scattering leads to a change in critical coupling conditions. When $\kappa_{\rm{out}} = 870~\rm{MHz}$, the transmission has a minimum value of  0.0055. (b) Cavity transmission spectra with different states of ion spin.}
\label{fig:photon gate}
\end{figure} 
     
\begin{figure}
\includegraphics[width=0.45\textwidth]{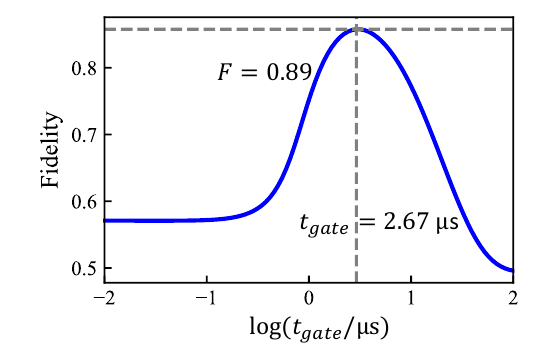}
\caption{Fidelity of the photon scattering gate as a function of the gate time $t_{\rm{gate}}$. The simulation is under \((\kappa, g, \gamma) = (50~\mathrm{MHz}, 20~\mathrm{MHz}, 10~\mathrm{kHz})\).}
\label{fig:F}
\end{figure} 

\section{\label{sec:level5}Photon gate }
This section examines the potential of coupled systems to achieve high-fidelity entanglement between flying and local qubits. Spin-photon interactions can be classified into four types: spin-photon emission, spin-photon gates, 
spin-photon absorption, and spin-photon projectors~\cite{beukers2024remote}.
Deterministic spin-photon gates and spin-photon absorption require strong coupling between the emitter and the cavity. In contrast, spin-photon emission and spin-photon projectors can operate effectively in the 'bad cavity' regime. Spin-photon gates or projectors can be implemented by utilizing the conditional reflection of the cavity's phase or amplitude. This serves as our primary method. To achieve this, it is essential to verify the transmission spectroscopy of the emitter-cavity system.

Considering the traveling wave solution of the WGM cavity, there is degeneracy in the propagation direction when the azimuthal mode number is the same. These two modes are called CW (clockwise) and CCW (counterclockwise) modes. However, defects within the cavity can cause scattering, leading to coupling between the CW and CCW modes. The effect of backscattering on ion spectra has been studied in quantum dot systems. ~\cite{srinivasan2007mode}. 

In light of the simplicity of the cavity modes, it is possible to express the Hamiltonian of the system as a function of three levels:
\begin{equation}
    H = H_0 + H_1
\end{equation}

\begin{equation}
\begin{aligned}
H_{0}&=\Delta_{cl}\hat{a}_{\mathrm{CW}}^{\dagger}\hat{a}_{\mathrm{CW}}+\Delta_{cl}\hat{a}_{\mathrm{CCW}}^{\dagger}\hat{a}_{\mathrm{CCW}} -\beta\hat{a}_{\mathrm{CW}}^{\dagger}\hat{a}_{\mathrm{CCW}}\\&-\beta^*\hat{a}_{\mathrm{CCW}}^\dagger\hat{a}_{\mathrm{CW}}+(E\hat{a}_{\mathrm{CW}}^\dagger+E^*\hat{a}_{\mathrm{CW}})
\end{aligned}
\label{eq:Hamiltonian of transmission0}
\end{equation}
 
\begin{equation}
\begin{aligned}
H_{1}&=\Delta_{al}\hat{\sigma}_{+}\hat{\sigma}_{+}g(\hat{a}_{\mathrm{CW}}^{\dagger}\hat{\sigma}_{-}+\hat{a}_{\mathrm{CW}}\hat{\sigma}_{+})\\&+g(\hat{a}_{\mathrm{CCW}}^\dagger\hat{\sigma}_-+\hat{a}_{\mathrm{CCW}}\hat{\sigma}_+)
\end{aligned}
\label{eq:Hamiltonian of transmission1}
\end{equation}
$\Delta_{cl} = \omega_c - \omega_l$, $\Delta_{al} = \omega_a - \omega_l$, and $\hat{a}_{\mathrm{CW}} (\hat{a}_{\mathrm{CCW}} )$ represent the annihilation operator of the CW (CCW) mode photon,  $\beta$ is the backscattering coupling strength with values around 1 GHz to 10 GHz. 
$E = \sqrt{\kappa_{\rm{out}} P_{in}}$ is the driving strength, and $\kappa_{\rm{out}}$ is the coupling strength between fiber and cavity.
For model simplicity, the effect of the phase is not considered.
Due to backscattering coupling, the cavity modes will be split by $\pm \beta$. This results in the effective coupling strength becoming $\sqrt{2}$ times the original value~\cite{srinivasan2007mode}. Backscattering leads to a change in critical coupling conditions. When $\kappa_{\rm{out}} = 870~\rm{MHz}$, the transmission has a minimum value of  0.0055, as shown in Fig.\ref{fig:photon gate}(a).
When $\beta \gg (\kappa ~,\gamma)$, the ions within the cavity are coupled to the $\omega - \beta$ mode. 

After the ion is prepared in the $\ket{\downarrow}$ state, the cavity and the atom are detuned, effectively making the cavity empty. The transmission spectrum of this system is determined using input-output theory~\cite{B1}:
\begin{equation}
    T_1 = \left\lvert t_1 \right\rvert ^2  = \left\lvert 1 - \frac{ \kappa_{\rm{out}} (i\Delta_{cl} + \kappa_{\rm{all}} /2)}{(i\Delta_{cl} + \kappa_{\rm{all}} /2)^2 + \beta^2}   \right\rvert ^2
    \label{eq:reflectance1}
\end{equation}
$C = \frac{4g^2}{\kappa_{\rm{\rm{all}}}\gamma}$ and we defined $\eta_{\rm{out}} = \kappa_{\rm{\rm{out}}} / (\kappa + \kappa_{\rm{\rm{out}}}) $, which means the ratio of the optical cavity decay through the fiber.
Due to the coupling of backscattering, the case of critical coupling is not reached when $\kappa_{\rm{\rm{out}}} = \kappa$. Correspondingly, the transmittance of the ion when it is prepared to the $\ket{\uparrow}$ state can be written as:
\begin{equation}
    T_2 =\left\lvert t_2 \right\rvert ^2 =\left\lvert 1-\frac{\kappa_{\rm{out}}(i\Delta_{cl}+\gamma / 2)}{2(i\Delta_{al}+\kappa_{\rm{all}}/ 2)(i\Delta+\gamma / 2)+4g^2 } \right\rvert ^2
    \label{eq:reflectance2}
\end{equation}
A resonant impedance mismatch resulting from vacuum Rabi splitting leads to cavity transparency. By monitoring the change in transmittance across different ionic states, the time-bin entanglement of photons is created~\cite{knaut2024entanglement}.

Initially, the ion is prepared in a superposition state using $\pi/2$ pulses. 
The time-bin photons are injected, and as the early pulse reaches the cavity, a $\pi$ pulse is applied to the ion, 
swapping its state. This establishes entanglement between a flying qubit and a local qubit, which is further used to entangle 
two distant local qubits. By guiding the photon into the next cavity and performing a Bell measurement on the photon, 
the entanglement between the two ions is confirmed. In this setup, partial cavity transmission introduces uncertainty when the ion is in the $\ket{\uparrow}$ state. The fidelity of the entanglement between flying and local qubits is expressed as:
\begin{equation}
    F = \frac{\left\lvert  \int_{}^{}  \,d\omega~ t_1(\omega)f(\omega)  \right\rvert ^2 (1 + \exp(-t_{g}/T_{d})) }{2(\left\lvert \int_{}^{}  \,d\omega~ t_1(\omega)f(\omega)  \right\rvert^2 + \left\lvert \int_{}^{}\,d\omega ~t_2(\omega)f(\omega)  \right\rvert^2)} 
    \label{eq:fidelity}
\end{equation}
where $f(\omega)$  is the envelope of the photon pulse spectrum. $T_{d}$ represents the qubit decoherence time, and $t_g$ is the gate time. 
The system's fidelity depends on the transmission spectrum's contrast and the photon pulses' duration. 
In the time domain, a short pulse duration prevents the ions within the cavity from reaching a stable state, 
making the cavity appear empty regardless of the ion state. 
In contrast, a longer pulse duration increases decoherence in the qubits.

Optimizing the transmission spectrum can improve the system's fidelity. The primary factor influencing $t_1$ is the system's coupling strength.
When the coupling strength is significantly greater than the cavity decay, $t_1$ approaches a value of 1 within the photon spectral range.
Conversely, $t_2$ is predominantly influenced by the external coupling efficiency. Transmission of resonance photons is zero only at critical coupling regions, ensuring high fidelity. 

In the actual system, a \(\mathrm{Yb^{3+}}\) ion doped in \(\mathrm{LiNO_3}\) has a coherence 
time \(T_{d} = 9.5 ~\rm{\mu s}\)~\cite{chiossi2022photon}, with coupling parameters
 \((\kappa, g, \gamma) = (50~\mathrm{MHz}, 20~\mathrm{MHz}, 1~\mathrm{kHz})\). 
Optical transition dephasing is neglected here. We begin by examining the case of critical coupling and model the transmission intensity 
across varying coupling strengths, finding the minimum transmission at 
 \(\kappa = 870~\mathrm{MHz}\).
The transmission function at  a steady state, evaluated at the resonance frequency according to 
Eq.~(\ref{eq:reflectance1}) and (\ref{eq:reflectance2}), is shown in FIG.~\ref{fig:photon gate}(b).

Given that \(\kappa \gg g\), the system operates in the 'bad cavity' regime. However, 
the splitting between \(\mathrm{Yb}\) ion qubits remains larger than the decay rate, 
ensuring well-defined qubit states. By considering a Gaussian pulse of time-bin photons 
with intervals greater than the qubit gate time, 
we derive a relationship between fidelity and pulse duration, as shown in FIG.~\ref{fig:F}. For a pulse duration 
\(t_{\rm{gate}}= 2.67~\rm{\mu s}\), the fidelity reaches a maximum of \(F = 0.889\). 
The primary limitation to this fidelity is the system's coupling strength, 
which could be increased five-fold through parametric driving, potentially yielding a 
fidelity of \(F = 0.95\) at \(t_{\rm{gate}}= 1.4~ \rm{\mu s}\).
The maximum fidelity depends on the decoherence time. Through dynamical decoupling, the decoherence 
time can be extended to milliseconds~\cite{heinze2014coherence}, allowing the maximum fidelity to reach approximately 0.99.

\section{\label{sec:level6}Conclusion and discussion}

This study explores the feasibility of achieving strong coupling between REIs and optical cavities, 
a critical component for advancing quantum networking and quantum information systems. 
Through theoretical and numerical analysis, we identify that WGM microcavities provide an optimal platform for strong coupling due to their unique ability to combine high-quality factors with small mode volumes.

Furthermore, our investigation shows that parametric amplification, particularly with $\rm{Yb^{3+}}$ ions doped in the lithium niobate, is promising for enhancing interaction strength in such systems. This approach effectively amplifies coupling and extends the potential for reliable photon-ion quantum state transfers within the cavity, which is crucial for applications such as photon gates and qubit entanglement.

In addition to examining the dynamics in the strong coupling region, we propose a framework for quantum networks based on photon scattering in cQED systems, leveraging the enhanced coupling. 
Our findings deepen the understanding of emitter-cavity dynamics and lay the 
groundwork for future experimental realizations in solid-state quantum devices.

\nocite{*}

\begin{acknowledgments}
This work is supported by the National Natural Science Foundation of China (Grants No.\ 12274400, 92265204, T2325023),  the Innovation Program for Quantum Science and Technology (Grants No.\ 2021ZD0302200), the China Postdoctoral Science Foundation (Grants No.\ 2024M763128), the Postdoctoral Fellowship Program of CPSF (Grants No.\ GZC20241652) and the Fundamental Research Funds for the Central Universities (Grants No.\ WK2030000076).
\end{acknowledgments}


%

\end{document}